\documentclass[a4paper,12pt]{article}
\usepackage[pctex32]{graphics}
\usepackage{amssymb,amsmath}

\begin{document}
\topmargin 0pt
\oddsidemargin 0mm
\newcommand{\be}{\begin{equation}}
\newcommand{\ee}{\end{equation}}
\newcommand{\ba}{\begin{eqnarray}}
\newcommand{\ea}{\end{eqnarray}}
\newcommand{\fr}{\frac}
\renewcommand{\thefootnote}{\fnsymbol{footnote}}

\begin{titlepage}

\begin{center}
{\Large \bf  Quasinormal frequencies and thermodynamic quantities
for the Lifshitz black holes }

\vskip .6cm
{\large   Yun Soo Myung$^{a}$ \footnote{e-mail
 address: ysmyung@inje.ac.kr} and Taeyoon Moon$^{b}$ \footnote{e-mail address: tymoon@sogang.ac.kr}}
 \\
\vspace{10mm} {{${}^{a}${\em Institute of Basic Science and School
of Computer Aided Science,  Inje University\\ Gimhae 621-749, Korea}
 \\
${}^{b}${\em Center for Quantum Space-time, Sogang University,
Seoul, 121-742, Korea}}} \vspace{5mm} \vskip .6cm

\end{center}

\begin{center}

\underline{Abstract}
\end{center}
We find the connection between thermodynamic quantities and
quasinormal frequencies in Lifshitz black holes.  It is shown that
the globally stable Lifshitz black holes have pure imaginary
quasinormal frequencies. We also show that by employing the
Maggiore's method,  both the horizon area and the entropy can be
quantized for these black holes.

\vspace{5mm}


\vskip 0.8cm


\thispagestyle{empty}
\end{titlepage}

\newpage
\section{Introduction}
The Lifshitz  black
holes~\cite{CFT-4,L-1,AL-3,L-2,L-4,L-3,L-5,Maeda2011} have received
considerable attentions since these may provide a model of
generalizing AdS/CFT correspondence to non-relativistic condensed
matter physics as the Lif/CFT
correspondence~\cite{CFT-1,CFT-2,CFT-3}.  Although their asymptotic
spacetimes are known to be Lifshitz, it is a non-trivial task to
find an analytic solution. One of the known solutions is  a
four-dimensional topological black hole which is asymptotically
Lifshitz with the dynamical exponent $z=2$~\cite{Mann}.  Analytic
black hole solutions with  planar horizon were  found in the
Einstein-scalar-massive vector theory~\cite{bm} and in the
Einstein-scalar-Maxwell theory~\cite{tay}. Another analytic solution
has  been recently found in the Lovelock gravity~\cite{MTr}. The
$z=3$ Lifshitz black hole~\cite{z3} was derived from the new massive
gravity (NMG)~\cite{bht}. Numerical solutions and thermodynamic
property of Lifshitz black hole  were  explored in~\cite{BBP,AL-2}.

A thermodynamic study  is important to understand  the Lifshitz
black hole  because heat capacity and free energy determine the
global stability of the Lifshitz black hole. A positive (negative)
heat capacity imply thermally stable (unstable) black hole, while a
positive (negative) free energy means unfavorable (favorable)
configuration in given ensemble. Hence, {\it a black hole with
positive heat capacity and negative free energy is considered as  a
globally stable black hole (GSBH)}.  However,  the thermodynamic
study on Lifshitz black holes was limited because it was difficult
to compute their conserved quantities in Lifshitz spacetimes.
Recently, there was a progress on computation of mass and related
thermodynamic quantities by using the ADT method~\cite{DS-1,DS-2}
and the Euclidean action approach~\cite{GTT}. Concerning the mass of
3D Lifshitz black hole, there is an apparent  discrepancy between
${\cal M}=\frac{7r_+^4}{8G_3 \ell^4}$ obtained from the ADT
method~\cite{DS-1} and ${\cal M}=\frac{r_+^4}{4G_3 \ell^4}$ from
other methods~\cite{GTT,HT,MKP}.  Phase transitions between Lifshitz
black holes and other configurations were investigated by using
on-shell and off-shell free energies~\cite{myung}.

On the other hand, quasinormal modes (QNMs) of a perturbed field
contain important information about the black hole. Their
quasinormal frequencies (QNFs)  are  given by
$\omega=\omega_R-i\omega_I$ whose real part represents the
perturbation oscillation and whose imaginary part denotes the rate
at which  this oscillation is damped, because of the presence of
black hole horizon.   In this sense, one requires $\omega_I>0$ which
is consistent with the stability condition of the black hole. Since
in asymptotically AdS (Lifshitz) spacetimes, spacelike infinity acts
like a reflecting boundary, Dirichlet, Neumann, or mixed boundary
condition have to be imposed there. The QNFs could be obtained by
solving the Klein-Gordon equation for a minimally coupled scalar by
imposing the boundary conditions: ingoing mode near the horizon and
Dirichlet  condition at infinity.

Importantly, all known QNFs~\cite{COP,GSV,GGLOR} have no  real part,
which implies that they are purely imaginary. If the purely
imaginary frequency ($\omega=-i \omega_I,~\omega_I>0$) represents an
interesting feature of the  Lifshitz black hole, it is very curious
to explore its connection to thermodynamic property. We note that
the imaginary part of QNFs involves the temperature of black hole
and a GSBH may provide an analytic form of QNFs.

 According to the  Hod's
conjecture~\cite{Hod}, the asymptotic QNFs is related to the
quantized black hole area. Identifying the vibrational frequency
$\omega(E)$ with the real part $\omega_R$, it leads to an  area
quantization of  $\Delta A_n=4\ln[3]\ell^2_{p}$ which is not
universal for all black holes. For a large damped case,
Maggiore~\cite{Mag} has proposed that the identification of
$\omega(E)$ with the imaginary part $\omega_I$ might lead to the
Bekenstein universal quantization of $\Delta
A_n=8\pi\ell^2_{p}$~\cite{Bek}. Hence, the analytic computation of
QNFs is  crucial for extracting  an important  information  on
Lifshitz black holes obtained from different gravitational theories.

In this work, we investigate  Lifshitz black holes by exploring  the
connection between  thermodynamic property and  quasinormal
frequencies. It is shown that globally stable Lifshitz black holes
provide purely imaginary quasinormal frequencies. We also find that
by using the Maggiore's method, the horizon area and entropy can be
equally spaced for these black holes.

 The organization of our
work is as follows. In section 2, we study the 3D Lifshitz black
hole by exploring a connection between its thermodynamic quantities
and QNFs, where QNFs are already known in \cite{COP}. We investigate
the 2D Lifshitz black hole by obtaining its purely imaginary QNFs in
section 3. In section 4, we study two 4D Lifshitz black holes: one
obtained from the Einstein-scalar-massive vector theory and the
other from the Einstein-scalar-Mawxwll theory. QNFs of the former
black hole were found in \cite{GSV,GGLOR} when replacing a radial
coordinate $r$ by $1/r$, while we obtain newly QNFs of the latter.
Finally, we find the area spectrum of two 4D Lifshitz black holes
after reviewing the 3D Lifshitz black hole in \cite{COP}.

\section{3D Lifshitz black hole}

The NMG ~\cite{bht} composed of the Einstein gravity with a
cosmological constant $\Lambda$ and higher-order curvature terms is
given by
\begin{eqnarray}
\label{NMGAct}
 S^{(3)}_{NMG} &=&-\Big[ S^{(3)}_{EH}+S^{(3)}_{HC}\Big], \\
\label{NMGAct2} S^{(3)}_{EH} &=& \frac{1}{16\pi G_3} \int d^3x \sqrt{-\cal{G}}~ ({\cal R}-2\Lambda),\\
\label{NMGAct3} S^{(3)}_{HC} &=& -\frac{1}{16\pi G_3\tilde{m}^2}
\int d^3x
            \sqrt{-\cal{G}}~\left({\cal R}_{\mu\nu}{\cal R}^{\mu\nu}-\frac{3}{8}{\cal R}^2\right),
\end{eqnarray}
where $G_3$ is a three-dimensional Newton constant and $\tilde{m}^2$
a parameter with mass dimension 2.  We  mention that to avoid
negative mass and entropy, it is necessary to take ``$-$" sign in
the front of $[ S^{(3)}_{EH}+S^{(3)}_{HC}]$.  The field equation is
given by \be {\cal R}_{\mu\nu}-\frac{1}{2}g_{\mu\nu}{\cal R}+\Lambda
g_{\mu\nu}-\frac{1}{2\tilde{m}^2}K_{\mu\nu}=0,\ee where
\begin{eqnarray}
  K_{\mu\nu}&=&2\square {\cal R}_{\mu\nu}-\frac{1}{2}\nabla_\mu \nabla_\nu {\cal R}-\frac{1}{2}\square{\cal R}g_{\mu\nu}\nonumber\\
        &+&4{\cal R}_{\mu\nu\rho\sigma}{\cal R}^{\rho\sigma} -\frac{3}{2}{\cal R}{\cal R}_{\mu\nu}-{\cal R}_{\rho\sigma}{\cal R}^{\rho\sigma}g_{\mu\nu}
         +\frac{3}{8}{\cal R}^2g_{\mu\nu}.
\end{eqnarray}
We have to choose
 $\tilde{m}^2=-\frac{1}{2\ell^2}$ and
$\Lambda=-\frac{13}{2\ell^2}$ to obtain the $z=3$  Lifshitz black
hole solution. Here  $\ell$  the curvature radius of Lifshitz
spacetimes.   Explicitly, we find the Lifshitz black hole
solution~\cite{z3} as
 \be
\label{3dmetric}
  ds^2_{\rm 3D}=g_{\mu\nu}dx^\mu dx^\nu=-\left(\frac{r^2}{\ell^2}\right)^z\left(1-\frac{M\ell^2}{r^2}\right)dt^2
   +\frac{dr^2}{\frac{r^2}{\ell^2}-M}+r^2d\phi^2,
\end{equation}
where $M$ is an integration constant related to the the  mass of
black hole. From the condition of $g^{rr}=0$, the event horizon is
determined to be  $r=r_+=\ell \sqrt{M}$.  This line element is
invariant under the anisotropic scaling of
\begin{equation}
t\to \lambda^zt,~~\phi \to \lambda \phi,~~ r\to
\frac{r}{\lambda},~~M\to \frac{M}{\lambda^2}.
\end{equation}
 For $z=1$, the ADM mass is determined to be
$M=\frac{r_+^2}{\ell^2}$, while for $z=3$, the ADM mass is not yet
fixed  completely.

All thermodynamic quantities  were driven by using the Euclidean
action approach. Its Hawking temperature ($T_H$), mass (${\cal M}$),
heat capacity ($C=\frac{d{\cal M}}{dT_H}$), Bekenstein-Hawking
entropy ($S_{BH}$), and Helmholtz free energy ($F={\cal
M}-T_HS_{BH}$) are given by \be \label{3Dther} T_H =
 \frac{r^3_+}{2\pi\ell^{4}},~{\cal
M}=\frac{r_+^4}{4G_3\ell^4},~C=\frac{4\pi r_+}{3G_3},~S_{BH} =
\frac{2\pi r_+}{G_3},~F=-\frac{3r_+^4}{4G_3\ell^4}. \ee At this
stage, we mention  a global structure of
 Lifshitz black hole. Its Penrose diagram is figured out  to be
 $\boxtimes$
 where a light-like curvature singularity is located at $r=0$ (top and bottom), while
 Lifshitz asymptote  is at $r=\infty$
(two sides).
 Generally, a black hole is increasing   by
absorbing radiations in the heat reservoir, while a black hole  is
decreasing  by Hawking radiation as evaporation process.  In
studying the phase transition, two important quantities are the heat
capacity $C$ which shows thermal stability (instability) for
$C>0(C<0)$ and free energy $F$ which indicates the global stability
for $F<0$.  For the case of positive heat capacity  and negative
free energy, we call it the globally stable black hole (GSBH). Here
it is observed from (\ref{3Dther})  that the Lifshitz black hole
belongs to the GSBH because of $C>0$ and $F<0$.

In order to make a connection to the QNMs, we consider the minimally
coupled scalar described by the Klein-Gordon equation
 \be \Big[\square_{\rm 3D}-m^2\Big] \varphi=0 \ee in the
background of Lifshitz black hole (\ref{3dmetric}) for $z=3$.
Decomposing $\varphi$ with $y=r_+/r$ as \be
\varphi(t,y,\phi)=R(y)e^{-i\omega t + i\kappa \phi},\ee the radial
equation takes the form \be \label{3Dre}
R''+\frac{y^2-3}{y(1-y^2)}R'+\frac{\ell^2}{1-y^2}\Bigg[\frac{\omega^2
y^4}{M^3(1-y^2)}-\frac{m^2}{y^2}-\frac{\kappa^2}{M\ell^2}\Bigg]R=0.
\ee Here we note that $r\in[r_+,\infty)$ is mapped inversely  to $
y\in [1,0)$. The solution to (\ref{3Dre}) is given by the confluent
Heun (HeunC) functions as \ba R(y)&=&C_1
y^{2+\alpha}(1-y^2)^{\frac{\beta}{2}} {\rm
HeunC}\Big[0,\alpha,\beta,-\frac{\beta^2}{4},\frac{\alpha^2}{4}+\frac{\kappa^2}{4M};y^2\Big]
\nonumber \\
&+& C_2 y^{2-\alpha}(1-y^2)^{\frac{\beta}{2}} {\rm
HeunC}\Big[0,-\alpha,\beta,-\frac{\beta^2}{4},\frac{\alpha^2}{4}+\frac{\kappa^2}{4M};y^2\Big],
\label{1sol11}\ea where $C_{1}$ and $C_{2}$ are arbitrary constants
and \be \alpha=2\sqrt{1+\frac{m^2\ell^2}{4}}>2,~~\beta=-i
\frac{\omega \ell}{M^{3/2}}=-i \frac{\omega}{2\pi T_H}. \ee
Requiring the Dirichlet condition at infinity ($y=0$) leads to
$C_2=0$ because of $2-\alpha<0$. In order to impose the ingoing mode
at horizon, one uses the connection formula \cite{Kwon}:
\begin{eqnarray}\label{connect}
&&{\rm HeunC}\Big[0,\alpha,\beta,\gamma,\delta;z\Big]
=\frac{\Gamma(\alpha+1)\Gamma(-\beta)}{\Gamma(1-\beta+K)\Gamma(\alpha-K)}
{\rm
HeunC}\Big[0,\beta,\alpha,-\gamma,\gamma+\delta;1-z\Big]\nonumber\\
&&\hspace*{3.5em}+~\frac{\Gamma(\alpha+1)\Gamma(\beta)}{\Gamma(1+\beta+S)\Gamma(\alpha-S)}
(1-z)^{-\beta} {\rm
HeunC}\Big[0,-\beta,\alpha,-\gamma,\gamma+\delta;1-z\Big],
\end{eqnarray}
where $K$ and $S$ are determined  by solving two algebraic equations
\begin{eqnarray}
K^2+(1-\alpha-\beta)K-\alpha-\beta-\epsilon+\frac{\gamma}{2}~=~0,
\nonumber\\
S^2+(1-\alpha+\beta)S-\alpha-\alpha\beta-\epsilon+\frac{\gamma}{2}
~=~0\nonumber
\end{eqnarray}
with $\epsilon=[1-(\alpha+1)(\beta+1)]/2-\delta$. Near the horizon
($y\to 1$), using (\ref{connect}),  the solution (\ref{1sol11})
 can be written by \be
R_{y\to1} =C_1\Big[ \xi_1(1-y^2)^{\frac{\beta}{2}}
+\xi_2(1-y^2)^{-\frac{\beta}{2}}\Big], \label{3rsol}\ee where
$\xi_{1}$ and $\xi_{2}$ are given by
\begin{eqnarray}
\xi_1=\frac{\Gamma(1+\alpha)\Gamma(-\beta)}
{\Gamma(\alpha-K)\Gamma(1-\beta+K)},
~~\xi_2=\frac{\Gamma(1+\alpha)\Gamma(\beta)}
{\Gamma(\alpha-S)\Gamma(1+\beta+S)}.
\end{eqnarray}
 For the real $\omega=\omega_R$ in
(\ref{3rsol}),  the former (latter) correspond to ingoing mode
$\longrightarrow \mid_{y=1}$ (outgoing mode $\longleftarrow
\mid_{y=1}$) because  the scalar field $\varphi$ behaves as
\begin{eqnarray}
\varphi&=& \xi_1e^{-i\omega
t}(1-y^2)^{\frac{\beta}{2}}+\xi_2e^{-i\omega
t}(1-y^2)^{-\frac{\beta}{2}} \nonumber \\
&\sim&\xi_1e^{-i\omega[t+\frac{1}{2\pi T_H}\ln(1-y)]}
+\xi_2e^{-i\omega[t-\frac{1}{2\pi T_H}\ln(1-y)]}
\end{eqnarray}
near the horizon. Hence,
 to obtain the ingoing mode at the horizon, one  imposes
 $\xi_2=0$ which implies that  $\Gamma(\alpha-S) \to \infty$~  or ~
 $\Gamma(1+\beta+S) \to \infty$.
This could  be done  by requiring \be \alpha-S=-n,~~~1+\beta
+S=-n,~n=0,1,2,\cdots\ee which lead to the same expression of
quasinormal frequency.

 Finally, one recovers  the purely imaginary
frequency~\cite{COP} \be \frac{\omega_{\pm}}{4\pi T_H }=-i
\Bigg[-1-2n -(4+m^2\ell^2)^{1/2}\pm\Big(7+\frac{3m^2\ell^2}{2}
+\frac{\kappa^2}{2M}+(3+6n)(4+m^2\ell^2)^{1/2}+6n(n+1)\Big)^{1/2}\Bigg].
\label{quasi3}\ee Its asymptotic frequency takes the form \be
\label{3dasym} \frac{\omega_+^{\infty}}{4 \pi T_H}=-i(\sqrt{6}-2) n,
\ee when choosing ``$\omega_+$'' because selecting ``$\omega_-$"
leads to  unstable quasinormal modes. At this stage, we note that
the expression (\ref{quasi3}) has already  appeared in~\cite{COP}.
However, it seems that in deriving this expression, they have made
two mistakes: one was to choose $C_1=0$, instead of $C_2=0$, and the
other was  to use a wrong connection formula, instead of a correct
one (\ref{connect}). Two mistakes happen to provide a correct
expression (\ref{quasi3}). In comparison to the BTZ quasinormal
frequency \be \omega^{BTZ}_{\pm}=\pm \frac{\kappa}{\ell}-i4\pi
T_{H}\Big[n+\frac{1}{2}(1+\sqrt{1+m^2\ell^2})\Big],\ee   the
imaginary frequency involves the angular quantum number $\kappa$ in
the Lifshitz black hole.

Consequently,  we confirm that the QNFs of 3D Lifshitz black hole,
which is a GSBH, are  purely imaginary. We check the connection
between purely imaginary quasinormal frequency and positive heat
capacity and negative free energy for the 3D Lifshitz black hole.

\section{2D Lifshitz black hole}
It is necessary to study a lower dimensional Lifshitz black hole
because the exact computation of QNFs  is available for the lower
dimensional gravitational system. The only known 2D Lifshitz black
hole could be obtained  when applying  the Achucarro-Ortiz (AO)
dimensional reduction to the NMG action~\cite{Achucarro}
\begin{eqnarray}
ds_{(3)}^2=g_{ij}dx^{i}dx^{j}+\ell^2\Phi^2d\theta^2
\end{eqnarray}
with the dilaton $\Phi$.  Integrating over $\theta$ on $S^1$, the
action (\ref{NMGAct}) reduces to the 2D effective dilaton action
as~\cite{MKP} \be S_{NMG} =-\Big[ S_{EH}+S_{HC}\Big], \ee
 where
\begin{eqnarray}
 S_{EH} &=& \ell \int d^2x \sqrt{-g}~ \Phi({R_{(2)}}-2\Lambda),\nonumber\\
 S_{HC} &=& -\frac{\ell}{2\tilde{m}^2} \int d^2x
            \sqrt{-g}~\Phi\Bigg[\frac{1}{4}R_{(2)}^2+\frac{1}{\Phi}R_{(2)}\nabla^2\Phi
            +\frac{2}{\Phi^2}\nabla_{i}\nabla_{j}\Phi\nabla^{i}\nabla^{j}\Phi
            -\frac{1}{\Phi^2}(\nabla^2\Phi)^2\Bigg].\nonumber
\end{eqnarray}
$S_{HC}$ contains fourth-order derivatives as the dilatonic kinetic
term. It turned out that for $\Phi=r/\ell$ and $z=3$, equations of
motion for 2D metric tensor $g^{ij}$ and dilaton $\Phi$ admit the 2D
Lifshitz black hole solution
\begin{eqnarray}\label{2dmetric}
ds^2_{\rm 2D}=g_{ij}dx^i
dx^j=-\left(\frac{r^2}{\ell^2}\right)^3\left(1-\frac{M\ell^2}{r^2}\right)dt^2
   +\frac{dr^2}{\frac{r^2}{\ell^2}-M}.
\end{eqnarray}
All thermodynamic quantities of 2D Lifshitz black hole are the same
as (\ref{3Dther})  of   the 3D Lifshitz black hole  because the
AO-dimensional reduction preserves all thermodynamic properties of
3D Lifshitz black hole. Hence, the 2D Lifshitz black hole is also a
GSBH.

 Now we introduce a minimally coupled
scalar equation \be \Big[\square_{\rm 2D}-m^2\Big] \psi=0 \ee in the
background of 2D Lifshitz black hole (\ref{2dmetric}) to find the
QNMs. Decomposing $\psi$ with $y=\ell \sqrt{M}/r=r_+/r$ as \be
\psi(t,\rho)=\rho(y)e^{-i\omega t},\ee the radial equation becomes
\be
\rho''-\frac{2}{y(1-y^2)}\rho'+\frac{\ell^2}{1-y^2}\Bigg[\frac{\omega^2
y^4}{M^3(1-y^2)}-\frac{m^2}{y^2}\Bigg]\rho=0, \label{2deq}\ee which
is similar to the 3D radial equation (\ref{3Dre}) with $\kappa=0$
($s$-mode). Also we note a  coordinate mapping: $r\in[r_+,\infty)
\to y\in (0,1]$. Solving the equation (\ref{2deq}) leads to the
 solution which is  expressed in terms of  the   HeunC functions
as \ba \rho(y)&=&\tilde{C}_1
y^{\frac{3}{2}+\gamma}(1-y^2)^{\frac{\beta}{2}} {\rm
HeunC}\Big[0,\gamma,\beta,-\frac{\beta^2}{4},\frac{\gamma^2}{4}+\frac{3}{16};y^2\Big]
\nonumber \\
&+& \tilde{C}_2 y^{\frac{3}{2}-\gamma}(1-y^2)^{\frac{\beta}{2}} {\rm
HeunC}\Big[0,-\gamma,\beta,-\frac{\beta^2}{4},\frac{\gamma^2}{4}+\frac{3}{16};y^2\Big],
\label{2sol3}\ea where $\tilde{C}_{1}$ and $\tilde{C}_{2}$ are
arbitrary constants and \be
\gamma=\frac{3}{2}\sqrt{1+\frac{4m^2\ell^2}{9}}>\frac{3}{2},~~\beta=-i
\frac{\omega \ell}{M^{3/2}}=-i \frac{\omega}{2\pi T_H}. \ee We
observe  that at infinity $(y\to0)$, imposing the Dirichlet
condition leads to $\tilde{C}_2=0$.  In order to obtain  the ingoing
mode at horizon, we   use the connection formula (\ref{connect}).
Near the horizon ($y \to 1$), the solution (\ref{2sol3}) takes the
form \be \rho_{y\to1} =\tilde{C}_1\Big[
\tilde\xi_1(1-y^2)^{\frac{\beta}{2}}
+\tilde\xi_2(1-y^2)^{-\frac{\beta}{2}}\Big], \label{2rsol}\ee where
the coefficients $\tilde\xi_{1}$ and $\tilde\xi_{2}$ are given by
\begin{eqnarray}\label{2sol4}
\tilde\xi_1=\frac{\Gamma(1+\gamma)\Gamma(-\beta)}
{\Gamma(\gamma-\tilde K)\Gamma(1-\beta+\tilde K)},
~~\tilde\xi_2=\frac{\Gamma(1+\gamma)\Gamma(\beta)}
{\Gamma(\gamma-\tilde S)\Gamma(1+\beta+\tilde S)}.
\end{eqnarray}
In these expressions, $\tilde K$ and $\tilde S$ are determined by
solving two equations
\begin{eqnarray}
\tilde K^2+(1-\gamma-\beta)\tilde
K-\gamma-\beta-\tilde\epsilon+\frac{\tilde \delta}{2}~=~0,
\nonumber\\
\tilde S^2+(1-\gamma+\beta)\tilde S-\gamma-\gamma\beta-\tilde
\epsilon+\frac{\tilde \delta}{2} ~=~0\nonumber
\end{eqnarray}
with $\tilde \delta=-\beta^2/4$ and $\tilde
\epsilon=[1-(\gamma+1)(\beta+1)]/2-\gamma^2/4-3/16$.

 For the real $\omega=\omega_R$ in
(\ref{2rsol}),  the former (latter) correspond to ingoing mode
(outgoing mode) because near the horizon,  the scalar field $\psi$
takes the form
\begin{eqnarray}
\psi &=& \tilde \xi_1e^{-i\omega
t}(1-y^2)^{\frac{\beta}{2}}+\tilde\xi_2e^{-i\omega
t}(1-y^2)^{-\frac{\beta}{2}} \nonumber \\
&\sim& \tilde\xi_1 e^{-i\omega[t+\frac{1}{4 \pi T_H}\ln(1-y)]}
+\tilde\xi_2e^{-i\omega[t-\frac{1}{4 \pi T_H}\ln(1-y)]}.
\end{eqnarray}
 To obtain the ingoing mode at the horizon, one has to impose
 $\tilde\xi_2=0$ which means  that  $\Gamma(\gamma-\tilde S) \to \infty$~  or ~
 $\Gamma(1+\beta+\tilde S) \to \infty$.
This is achieved  by  requiring \be \gamma-\tilde S=-n,~~~1+\beta
+\tilde S=-n,~n=0,1,2,\cdots \ee which lead to the same QNFs.

Consequently, one obtains the purely imaginary frequency
\be\label{quasi2} \frac{\omega_\pm}{4\pi T_H }=-i \Bigg[-1-2n
-\Big(\frac{9}{4}+m^2\ell^2\Big)^{1/2}\pm\Big(\frac{19}{4}+\frac{3m^2\ell^2}{2}
+(3+6n)\Big(\frac{9}{4}+m^2\ell^2\Big)^{1/2}+6n(n+1)\Big)^{1/2}\Bigg].
\ee We note that its asymptotic frequency takes the same form as
that of 3D Lifshitz black hole  in (\ref{3dasym}) \be
\frac{\omega_+^{\infty}}{4 \pi T_H}=-i(\sqrt{6}-2) n \ee which shows
that 2D Lifshitz black hole is very similar to  3D Lifshitz black
hole. In comparison, we introduce  the QNFs for the  AdS$_2$ black
hole~\cite{COV} \be \label{aoq} \frac{\omega_{AO}}{2\pi
T_H}=-i\Bigg(n+\frac{1+\sqrt{1+4m^2\ell^2}}{2}\Bigg) \ee which is
purely imaginary.

As a result, it is shown that a scalar propagating in the 2D
Lifshitz spacetimes provides the purely imaginary quasinormal
frequency. We find the connection between purely imaginary
quasinormal frequency and positive heat capacity and negative free
energy   for  2D Lifshitz black hole.

\section{4D Lifshitz black holes}

Up to now, we have found  the connection between purely imaginary
quasinormal frequency and positive heat capacity and negative free
energy   for  2D and 3D Lifshitz black holes. In order to confirm
this suggested connection, it is necessary  to investigate 4D
Lifshitz black holes. Two known Lifshitz black holes were obtained
from the Einstein-scalar-massive vector theory and the
Einstein-scalar-Maxwell theory.

\subsection{Einstein-scalar-massive vector theory}
First we study the 4D1 Lifshitz black hole obtained from the
Einstein-scalar-massive vector theory~\cite{bm}
\begin{equation}
S_{\rm ESMV}=\frac{1}{16\pi G_{4}}\int
d^{4}x\sqrt{-g}[R-2\Lambda-\tilde{m}^2A_{\mu}A^{\mu}-2(e^{-2\phi}-1)
-\frac{1}{2}e^{-2\phi}F_{\mu\nu}F^{\mu\nu}],
\end{equation}
where \be \Lambda=-\frac{z^2+z+4}{2},~\tilde{m}^2=2z,~F=dA \ee with
$L^2=1$. The Lifshitz black hole solution for $z=2$ is given by \ba
ds^2_{\rm
4D1}&=&-r^{2z}f(r)dt^2+r^2\Big(dx_1^2+dx_2^2\Big)+\frac{dr^2}{f(r)r^2},
\nonumber
\\ e^{-2\phi}&=&1+\frac{r_+^2}{r^2},~A=\frac{f(r)r^2dt}{\sqrt{2}}
\label{4lif1}\ea with the metric function
\begin{eqnarray}
f(r)=1-\frac{r_+^2}{r^2}. \nonumber
\end{eqnarray}
It is important
to note that the above solution could be   obtained by replacing $r$
by $1/r$ in the original solution appeared in~\cite{CFT-4}.

Applying the Euclidean action approach to this theory, one finds
thermodynamic quantities \be \label{tdq1}T_{H}=\frac{r^2_+}{2\pi
},~M=\frac{V_2r_+^4}{16\pi G_4},~C=\frac{ V_2r_+^2}{4G_4
},~S_{BH}=\frac{ r_+^2V_2}{4G_4 },~F=-\frac{V_2r_+^4}{16\pi
 G_4 }, \ee
 where $V_2$ is the volume of the transverse directions.
The 4D1 Lifshitz black hole is a GSBH because of $C>0$ and $F<0$.
 We
note that the  thermodynamic quantities in (\ref{tdq1})  are
obtained by replacing $r_+$ by $1/r_+$ in the original expressions
as \be T_{H}=\frac{1}{2\pi r^2_+},~M=\frac{V_2}{16\pi
G_4r_+^4},~C=\frac{ V_2}{4G_4 r_+^2},~S_{BH}=\frac{ V_2}{4G_4
r_+^2},~F=-\frac{V_2}{16\pi
 G_4 r_+^4}. \ee
Here,  the limit  $r_+ \to 0$ of all thermodynamic quantities goes
to infinity which indicates that the original radial coordinate is
not appropriate for describing thermodynamics of  the Lifshitz black
hole. In addition, the computation of QNFs  has been performed by
using the original  coordinate appeared in~\cite{GSV}, showing four
types of QNFs: $\omega_{i},~i=1,2,3,4$.  This is the reason why we
use the new radial coordinate as the inverse of original coordinate.

 In order to compute QNFs for the $z=2$ Lifshitz black
hole (\ref{4lif1}),   we  consider a massive scalar field given by
\be \Big[\square_{\rm 4D1}-m^2\Big] \Psi_1=0.\label{4main}\ee
Assuming $\Psi_1=H_1(r)e^{-i\omega t}e^{-i(k_1x_1+k_2x_2)}$ and
introducing $y=r_+/r$, equation (\ref{4main}) becomes
\begin{eqnarray}\label{4eq1}
H_1''+\frac{y^2-3}{y(1-y^2)}H_1'+\frac{y^4\omega^2
-r_+^2(r_+^2m^2+k^2y^2)(1-y^2)}{r_+^4y^2(1-y^2)^2}H_1=0,
\end{eqnarray}
where $k^2=k_1^2+k_2^2$ and the prime ($'$) denotes the
differentiation with respect to $y$. It is found that the solution
to Eq.(\ref{4eq1}) is given by the hypergeometric functions
\begin{eqnarray}\label{4sol1}
H_1(y)&=&D_1
y^{\alpha}(1-y^2)^{\beta}~{}_{2}F_{1}\Big[a,b,c;y^2\Big]
\nonumber \\
&+&D_2
y^{4-\alpha}(1-y^2)^{\beta}~{}_{2}F_{1}\Big[a-c+1,b-c+1,2-c;y^2\Big],
\end{eqnarray}
where ${}_{2}F_{1}$ is the hypergeometric function, $D_{1}$  and
$D_{2}$ are arbitrary constants, and
\begin{eqnarray}
\alpha=2\Big[1+\sqrt{1+\frac{m^2}{4}}\Big]>4,~~~\beta=-\frac{i\omega}{4\pi
T_H}.
\end{eqnarray}
In the hypergeometric function, the arguments of $a,~b,$ and $c$ are
given by
\begin{eqnarray}
a&=&\beta+\frac{\alpha}{2}-\frac{1}{2}
-\frac{1}{2r_+^2}\sqrt{r_+^4-\omega^2-r_+^2k^2},\\
b&=&\beta+\frac{\alpha}{2}-\frac{1}{2}
+\frac{1}{2r_+^2}\sqrt{r_+^4-\omega^2-r_+^2k^2},\\
c&=&\alpha-1.
\end{eqnarray}
At infinity ($y\to0$), we see that $D_2$ of (\ref{4sol1}) should be
zero because it corresponds to non-normalizable mode ($4-\alpha<0$).
At the horizon ($y=1$), using the connection formula \cite{Abram}
which  connects $y^2$ to $1-y^2$  for the hypergeometric function,
the  solution (\ref{4sol1}) becomes
\begin{eqnarray}\label{4sol2}
H_1(y)&=&D_1 y^{\alpha}\Bigg[(1-y^2)^\beta
\frac{\Gamma(c)\Gamma(c-a-b)}{\Gamma(c-a)\Gamma(c-b)}
{}_{2}F_{1}\Big[a,b,a+b-c+1;1-y^2\Big]\nonumber\\
&+&(1-y^2)^{-\beta}
\frac{\Gamma(c)\Gamma(a+b-c)}{\Gamma(a)\Gamma(b)}
{}_{2}F_{1}\Big[c-a,c-b,c-a-b+1;1-y^2\big]\Bigg].
\end{eqnarray}
It is worth to note that at the horizon, the first term in
(\ref{4sol2}) corresponds to the ingoing mode, while the second
corresponds  to the outgoing mode. The ingoing mode  near the
horizon ($y\to1)$ is obtained  by imposing the condition
\begin{eqnarray}
\frac{\Gamma(c)\Gamma(a+b-c)}{\Gamma(a)\Gamma(b)}=0,
\end{eqnarray}
which implies that
\begin{eqnarray}\label{ab}
a=-n,~~~b=-n,~n=0,1,2,\cdots.
\end{eqnarray}

Consequently,  we find the purely imaginary quasinormal frequency
as
\begin{eqnarray} \label{4d1q}
\frac{\omega_{\rm 4D1}}{2\pi
T_H}&=&-i\Bigg[\frac{(2n+1)\Big((4n^2+4n-4-m^2)+\frac{k^2}{2\pi
T_H}\Big)}{2(4n^2+4n-3-m^2)} \nonumber \\
&+&~~~~~ \frac{\Big((4n^2+4n-2-m^2)-\frac{k^2}{2\pi
T_H}\Big)\sqrt{4+m^2}}{2(4n^2+4n-3-m^2)} \Bigg]
\end{eqnarray}
which is the same expression as those of  $\omega_2$ in~\cite{GSV}
and $\omega$ in~\cite{GGLOR}. The $n=0$ case leads to $\omega_I>0$,
which means that the 4D1 Lifshitz black hole are stable against the
external perturbations. Also,  we observe that in the large $n$
limit, $\omega_{\rm 4D1}$ becomes \be \label{4d1qi}
 \frac{\omega^\infty_{\rm 4D1}}{2\pi T_H}=-in,\ee
which will be used to derive the area quantization in section 5.

Finally, we confirm that a scalar propagating in the 4D1 Lifshitz
black hole has  purely imaginary quasinormal frequency. We find  the
connection between purely imaginary quasinormal frequency and
positive heat capacity and negative free energy for the 4D1 Lifshitz
black hole obtained from the Einstein-scalar-massive vector theory.

\subsection{Einstein-scalar-Maxwell theory}
We introduce the action of Einstein-scalar-Maxwell theory~\cite{tay}
\begin{equation}
S_{\rm ESM}=\frac{1}{16\pi G_{4}}\int
d^{4}x\sqrt{-g}[R-2\Lambda-\frac{1}{2}\partial_{\mu}\phi\partial^{\mu}\phi
-\frac{1}{4}e^{\tilde{\lambda}\phi}F_{\mu\nu}F^{\mu\nu}],
\end{equation}
where $\Lambda$ is the cosmological constant and two fields are a
massless scalar and a Maxwell field. It admits the 4D2 Lifshitz
black hole with  $z=2$ as solution to  equations of
motion~\cite{pang2}
\begin{eqnarray}
\label{2eq2} &
&ds^{2}_{\rm 4D2}=L^{2}\Big[-r^{2z}f(r)dt^{2}+\frac{dr^{2}}{r^{2}f(r)}+r^{2}\sum\limits^{2}_{i=1}
dx^{2}_{i}\Big],\nonumber\\
& &f(r)=1-\frac{r_{+}^{z+2}}{r^{z+2}},~~~e^{\tilde{\lambda}\phi}=\frac{1}{r^{4}},~~~\tilde{\lambda}^{2}=\frac{4}{z-1},\nonumber\\
&
&F_{rt}=qr^{z+1},~~~\Lambda=-\frac{(z+1)(z+2)}{2L^{2}},\nonumber\\
& &q^{2}=2L^{2}(z-1)(z+2),
\end{eqnarray}
where the event horizon is located at $r=r_{+}$. This line element
is invariant under the anisotropic scaling of $t\to \lambda^zt,x_i
\to \lambda x_i, r\to r/\lambda$, and $r_+\to r_+/\lambda$. It is
important to note from the last relation of (\ref{2eq2}) that the
charge $q$ is not an independent  hair because it is determined by
the curvature radius $L$ of Lifshitz black hole and its dynamical
exponent $z$,  which contrasts to the Reissner-Nordstr\"om-AdS black
hole. A similar case was found in the charged MTZ black
hole~\cite{cmtz,mp}

The temperature and Bekenstein-Hawking entropy are determined by
\begin{equation}
\label{2eq3} T^z_H=\Big[\frac{z+2}{4\pi}\Big]r^{z}_{+},~S_{\rm
BH}=\frac{L^{2}V_{2}}{4G_{4}}r^{2}_{+},
\end{equation}
where $V_{2}$ denotes the volume of  two-dimensional spatial
directions.  Mass, heat capacity, and Helmholtz free energy are
obtained by using Euclidean action approach as \be M^z=\frac{2L^2V_2
r_+^{z+2}}{16\pi
G_4},~C^z=\frac{dM^z}{dT^z_H}=\frac{2L^2V_2r_+^2}{4zG_4},~F^z=-\frac{zL^2V_2
r_+^{z+2}}{16\pi G_4}, \ee where $F^z$ is different from the Gibbs
free energy defined by $\tilde{F}^z=-\frac{L^2V_2 r_+^{z+2}}{16\pi
G_4}$. We stress that the 4D2 Lifshitz black hole is also a GSBH
because of $C^z>0$ and $F^z<0$.

 In order to compute QNFs for the  $z=2$ Lifshitz black
hole (\ref{2eq2}),  we consider a massive scalar field given by \be
\Big[\square_{\rm 4D2}-m^2\Big] \Psi_2=0.\label{4main2}\ee Assuming
$\Psi_2=H_2(r)e^{-i\omega t}e^{-i(k_1x_1+k_2x_2)}$ and introducing
$y=r_+/r$, equation (\ref{4main2}) becomes
\begin{eqnarray}\label{4eq2}
H_2''-\frac{3+y^4}{y(1-y^4)}H_2'+\frac{y^4\omega^2
-r_+^2(k^2y^2+r_+^2L^2m^2)(1-y^4)}{r_+^4y^2(1-y^4)^2}H_2=0,
\end{eqnarray}
where $k^2=k_1^2+k_2^2$ and the prime ($'$) denotes the
differentiation with respect to $y$. Importantly, the solution to
(\ref{4eq2}) is given by the general Heun (HeunG) function as
follows:
\begin{eqnarray}
H_2(y)&=&\tilde{D}_1
y^{2+\alpha}(1+y^2)^{-\frac{\beta}{2}}(1-y^2)^{\frac{\beta}{2}} {\rm
HeunG}\Big[-1,~-\gamma,~1+\frac{\alpha}{2},~1+\frac{\alpha}{2},~
1+\alpha,~1+\beta;~y^2\Big]
\nonumber \\
&&\hspace*{-4em} +\tilde{D}_2
y^{2-\alpha}(1+y^2)^{-\frac{\beta}{2}}(1-y^2)^{\frac{\beta}{2}} {\rm
HeunG}\Big[-1,~-\gamma+2\alpha\beta,~1-\frac{\alpha}{2},~
1-\frac{\alpha}{2},~1-\alpha,~1+\beta;~y^2\Big],\nonumber\\
\label{4sol4}
\end{eqnarray} where $\tilde{D}_{1}$ and  $\tilde{D}_{2}$ are arbitrary
constants, and
\begin{eqnarray}
&&\alpha=2\sqrt{1+\frac{m^2L^2}{4}}>2,~~~~~\beta=-i \frac{\omega
}{2\pi T_H},~~~~\gamma=(1+\alpha)\beta+\frac{k^2}{4\pi T_H}.
\end{eqnarray} At infinity ($y=0$), we see that $\tilde D_2$
 should be zero
because it corresponds to non-normalizable mode ($2-\alpha<0$). In
order to derive  the asymptotic form of the HeunG function near the
horizon ($y\to1$), we use the  formula ~\cite{book,HG}
\begin{eqnarray}\label{fomul}
&&{\rm HeunG}\Big[b_1,~ b_2,~ a_1,~ a_2,~ a_3,~ a_4;~ z\Big]\nonumber\\
&=&(1-z)^{1-a_4} {\rm HeunG}\Big[b_1,~ b_2-(a_4-1)a_3b_1,~
 a_2-a_4+1,~ a_1-a_4+1,~ a_3,~ 2-a_4;~ z\Big]\nonumber\\
 &&\nonumber\\
 &=&E_1~{\rm HeunG}\Big[1-b_1,~-b_2-a_1a_2,~a_1,~a_2,~1+a_1
 +a_2-a_3-a_4,~a_4;~1-z\Big]+\nonumber\\
 &&\hspace*{-1em}(1-z)^{a_3+a_4-a_1-a_2}E_2~{\rm HeunG}\Big[1-b_1,
 ~-b_2-a_1a_2-(a_3+a_4-a_1-a_2)(a_3+a_4-b_1a_3),\nonumber\\
 &&\hspace*{7em}~a_3+a_4-a_1,~a_3+a_4-a_2,
 ~1-a_1-a_2+a_3+a_4,~a_4;~1-z\Big],
\end{eqnarray}
where $E_1$ and $E_2$ are given by
\begin{eqnarray}
&&\hspace*{-2em}E_1={\rm HeunG}\Big[b_1,~ b_2,~ a_1,~ a_2,~ a_3,
~ a_4;~ 1\Big],\label{E12}\\
&&\hspace*{-2em}E_2={\rm HeunG}\Big[b_1,~
b_2-b_1a_3(a_3+a_4-a_1-a_2), ~ a_3+a_4-a_1,~ a_3+a_4-a_2,~ a_3,~
a_4;~ 1\Big].\label{E22}
\end{eqnarray}
Using the  formula (\ref{fomul}),  the solution (\ref{4sol4})  takes
the form near the horizon
\begin{eqnarray}\label{4solF}
H_2(y)=\tilde{D_1}y^{2+\alpha}(1+y^2)^{\frac{\beta}{2}}
\Big[~(1-y^2)^{-\frac{\beta}{2}}E_1+
(1-y^2)^{\frac{\beta}{2}}E_2\Big].
\end{eqnarray}
We note that  the first term in (\ref{4solF}) corresponds to the
outgoing mode,  while the second corresponds  to the ingoing mode.
To obtain  the ingoing mode for the solution (\ref{4solF}) near the
horizon ($y\to1)$,  we impose $E_1=0$ which implies
\begin{eqnarray}\label{4sol41}
{\rm HeunG}\Big[-1,~-\frac{k^2}{4r_+^2},~1+\frac{\alpha}{2}-\beta,
~1+\frac{\alpha}{2}-\beta,~1+\alpha,~1-\beta;~1\Big]=0.
\end{eqnarray}
At this stage, we have to  mention that a little bit of  general
Heun function and its connection formula is   known, in comparison
with  the HeunC function and hypergeometric function. Hence, it
seems to be a formidable task to obtain  QNFs from a general
condition of (\ref{4sol41}). Fortunately, we observe that for
$k^2=0$ ($s$-mode),  (\ref{4sol41}) reduces to a condition for the
hypergeometric function
\begin{eqnarray}\label{fh}
{}_{2}F_{1}\Big[\frac{1}{2}+\frac{\alpha}{4}-\frac{\beta}{2},
~\frac{1}{2}+\frac{\alpha}{4}-\frac{\beta}{2},
~1+\frac{\alpha}{2};~1\Big]=0
\end{eqnarray}
In deriving this condition, we  used the  reduction formula
\cite{diff}
\begin{eqnarray}
{\rm
HeunG}\Big[-1,~0,~a_1,~a_2,~a_3,~\frac{a_1+a_2-a_3+1}{2};~z\Big]
={}_{2}F_{1}\Big[\frac{a_1}{2},~\frac{a_2}{2},~\frac{1+a_3}{2},~z^2\Big].
\end{eqnarray}
Considering   a relation for the hypergeometric function
\begin{eqnarray}
{}_{2}F_{1}[c_1,~c_2,~c_3;~1]=\frac{\Gamma(c_3)\Gamma(c_3-c_1-c_2)}
{\Gamma(c_3-c_1)\Gamma(c_3-c_2)},
\end{eqnarray}
the condition of  ${}_{2}F_{1}[c_1,~c_2,~c_3;~1]=0$ implies that
\begin{eqnarray}\label{condf}
c_3-c_1=-n,~~~{\rm or}~~~c_3-c_2=-n,~n=0,1,2,\cdots.
\end{eqnarray}
 Applying the condition (\ref{condf}) to
(\ref{fh}) leads to  the purely imaginary  quasinormal mode as
\begin{eqnarray}\label{43q}
\frac{\omega_{k^2=0}}{4\pi T_H}=-i\Bigg[n+
\frac{1+\sqrt{1+\frac{m^2L^2}{4}}}{2}\Bigg],
\end{eqnarray}
which is similar to (\ref{aoq}) of the AdS$_2$ black hole. In the
large $n$ limit, the QNFs behave as
\begin{eqnarray}\label{43qi} \frac{\omega^\infty_{k^2=0}}{4\pi
T_H}=-i n,
\end{eqnarray}
which will be used to derive the area quantization of 4D2 Lifshitz
black hole.

Finally,   to confirm  the QNFs of $s$-mode, we start with  equation
(\ref{4eq2}) with $k^2=0$,
\begin{eqnarray}\label{42eq1}
\tilde{H}_2''-\frac{3+y^4}{y(1-y^4)}\tilde{H}_2'+\frac{(y^4\omega^2
-r_+^4L^2m^2)(1-y^4)}{r_+^4y^2(1-y^4)^2}\tilde{H}_2=0.
\end{eqnarray}
The solution to (\ref{42eq1})  is given by the hypergeometric
functions
\begin{eqnarray}\label{42sol1}
\tilde{H}_2(y)&=&d_1
y^{2-\alpha}(1-y^4)^{-\frac{\beta}{2}}~{}_{2}F_{1}
\Big[-\frac{\beta}{2}-\frac{\alpha}{4}+\frac{1}{2},~
-\frac{\beta}{2}-\frac{\alpha}{4}+\frac{1}{2},~1-\frac{\alpha}{2};~y^4\Big]
\nonumber\\&&\hspace*{2em} +~d_2
y^{2+\alpha}(1-y^4)^{-\frac{\beta}{2}}~
{}_{2}F_{1}\Big[-\frac{\beta}{2}+\frac{\alpha}{4}+\frac{1}{2},~
-\frac{\beta}{2}+\frac{\alpha}{4}+\frac{1}{2},~1+\frac{\alpha}{2};~y^4\Big]
\end{eqnarray}
 with $d_{1}$ and $d_2$ arbitrary constants.
At infinity ($y=0$), $d_1$  should be zero because of $2-\alpha<0$.
Near the horizon ($y\to1$), using the connection formula
\cite{Abram}, the solution (\ref{42sol1}) becomes
\begin{eqnarray}\label{42sol2}
\tilde{H}_2(y)=d_2
y^{2+\alpha}\Big[\zeta_1(1-y^4)^{-\frac{\beta}{2}}
+\zeta_2(1-y^4)^{\frac{\beta}{2}}\Big]
\end{eqnarray}
with the coefficients
\begin{eqnarray}
\zeta_1=\frac{\Gamma\Big(1+\frac{\alpha}{2}\Big)\Gamma\Big(\beta\Big)}
{\Gamma\Big(\frac{\alpha}{4}+\frac{\beta}{2}+
\frac{1}{2}\Big)\Gamma\Big(\frac{\alpha}{4}+\frac{\beta}{2}+
\frac{1}{2}\Big)},~~~
\zeta_2=\frac{\Gamma\Big(1+\frac{\alpha}{2}\Big)\Gamma\Big(-\beta\Big)}
{\Gamma\Big(\frac{\alpha}{4}-\frac{\beta}{2}+
\frac{1}{2}\Big)\Gamma\Big(\frac{\alpha}{4}-\frac{\beta}{2}+
\frac{1}{2}\Big)}.
\end{eqnarray}
Near the horizon the first term in (\ref{42sol2}) corresponds to the
outgoing mode,  while the second term corresponds  to the ingoing
mode.   To have the ingoing mode near the horizon, we impose
$\zeta_1=0$ to give
\begin{eqnarray}\label{ab3}
\frac{\alpha}{4}+\frac{\beta}{2}+\frac{1}{2}=-n, ~n=0,1,2,\cdots.
\end{eqnarray}
 The condition of (\ref{ab3}) provides us
purely imaginary  QNFs as
\begin{eqnarray}
\omega=-ir_+^2\Big[4n+2+\sqrt{4+L^2m^2}\Big],
\end{eqnarray}
which is exactly the same form  as in (\ref{43q}).

Consequently, we have shown that $s$-mode QNFs of 4D2 Lifshitz black
hole are  purely imaginary. Our question is `` Can QNFs of Lifshitz
black hole remain purely imaginary form even if $k^2\not=0$ ?". At
this stage, we could not  answer to this question. However, the
observation of QNFs of Lifshitz black holes suggests that $k^2$-term
includes as a term in $\omega_I$ with $\omega _R=0$.

\section{Area spectrum of Lifshitz black holes}
It was suggested  that black holes could  provide a test bed for any
proposed scheme for  quantum theory of gravity. In this direction,
Hod  has  combined the perturbation of  black holes with the quantum
mechanics and statistical physics  to derive the quantum of the
black hole area spectrum~\cite{Hod}.  For a  highly excited
Schwarzschild black hole,  Hod has used the real part $\omega_R$ of
QNFs  to obtain  the area quanta of  $\Delta A_n= 4\ln[3]l_p^2$.
However, it is not consistent with $\Delta A_n= 8\pi l_p^2$  which
was obtained  from the fact that the black hole area is
adiabatically invariant by Bekenstein~\cite{Bek}.   Kunstatter has
shown that the area spectrum is equally spaced for higher
dimensional Schwarzschild black holes when using the adiabatically
invariant integral~\cite{Kun}
 \be \label{adiabatic} I=\int \frac{dE}{\omega(E)} \to \int
\frac{dM}{\omega_R},\ee where $(E,\omega)$ are  (energy, vibrational
frequency) and $(M,\omega_R)$ are (black hole mass, real part of
QNFs). On later, Maggiore has proposed that a black hole perturbed
by external field is considered as a collection of damped harmonic
oscillators~\cite{Mag}. He has regarded
$\omega_0=\sqrt{\omega_R^2+\omega_I^2}$ as a physically proper
frequency and thus, $\omega_0= \omega_I$ was used to derive $\Delta
A_n= 8\pi l_p^2$  for highly excited QNFs of $\omega_I \gg \omega_R$
by considering the transition from $n $ to $n-1$.

Since  QNFs of Lifshitz black holes are known to be purely imaginary
and all thermodynamic quantities of Lifshitz black holes are known,
we could use these to derive area spectrum of Lifshitz black holes
by using Maggiore's method solely. In other words, the Hod's method
is inapplicable to extracting information on Lifshitz black holes.
Especially, we have their asymptotic frequencies which are  given by
(\ref{3dasym}) for 3D Lifshitz black hole, (\ref{4d1qi}) and
(\ref{43qi}) for 4D Lifshitz black holes.   We review how to derive
the quantization of horizon area  for the 3D Lifshitz black
hole~\cite{COP}.  We compute the adiabatic invariant $I$ \be I=\int
\frac{dM}{\omega_c}=\frac{r_+}{2G_3(\sqrt{6}-2)}, \ee where \be
\omega_c=i\Big[(\omega_+^\infty)_n-(\omega^\infty_+)_{n-1}\Big]=4\pi
T_H (\sqrt{6}-2)=2(\sqrt{6}-2)\frac{r_+^3}{\ell^4}. \ee Using the
Bohr-Sommerfeld quantization condition of $I \approx n \hbar$ and
considering the horizon area $A=2\pi r_+$, the quantized area
spectrum is given by \be A_n= 4 \pi G_3 (\sqrt{6}-2) n \hbar, \ee
which is not the universal area spectrum of $A_n^{u}=8\pi n \hbar $.
The entropy spectrum takes the form \be S^n_{BH}=\frac{A_n}{G_3}=4
\pi(\sqrt{6}-2) n \hbar. \ee

For 4D1 Lifshitz black hole, the adiabatic invariant $I$ is given by
\be I_{\rm 4D1}=\int \frac{dM}{\omega_c}=\frac{V_2r_+^2}{4\pi G_4},
\ee where \be \omega_c=i\Big[(\omega^{\infty}_{\rm
4D1})_n-(\omega^{\infty}_{\rm 4D1} )_{n-1}\Big]=2\pi T_H=r_+^2. \ee
Considering the horizon area $A=V_2r_+^2$, the quantized area
spectrum is given by \be A_n= 4 \pi G_4  n \hbar, \ee which is the
universal area spectrum of $A_n^{u}=8\pi n \hbar $ with $G_4=2$. In
addition, the entropy spectrum can be obtained by \be S^n_{\rm
4D1}=\frac{A_n}{4G_4}=\pi n \hbar. \ee

For other 4D2 Lifshitz black hole with $L^2=1$, the adiabatic
invariant $I$ takes the form \be I_{\rm 4D2}=\int
\frac{dM}{\omega_c}=\frac{V_2r_+^2}{8\pi G_4}, \ee where \be
\omega_c=i\Big[(\omega^{\infty}_{k^2=0})_n-(\omega^{\infty}_{k^2=0}
)_{n-1}\Big]=4\pi T_H=4r_+^2. \ee Considering the horizon area
$A=V_2r_+^2$, the quantized area spectrum is given by \be A_n= 8 \pi
G_4  n \hbar, \ee which is not the universal area spectrum of
$A_n^{u}=8\pi n \hbar $ with $G_4=2$. In this case one finds that
the entropy spectrum is given by \be S^n_{\rm
4D2}=\frac{A_n}{4G_4}=2 \pi n \hbar. \ee

\section{Discussions}
First of all,  the purely imaginary QNFs show that scalar
perturbation  has no considerable oscillation stage around the
Lifshitz black hole.  This implies that the equilibrium is stable
and thus, it is  difficult to deviate the black hole from its
equilibrium configuration.   If the Lif/CFT correspondence exists
really,  the thermalization timescale of the boundary conformal
field theory is given by $\tau=\frac{1}{4\pi T_H}$ which implies
that at high temperature, the field theory time scale is very small,
indicating that a perturbation  in the boundary conformal field
theory is not long-lived and it decreases exponentially to zero.

We regard the purely imaginary QNFs  as an interesting feature of
the Lifshitz black hole, in comparison  to the AdS black hole. In
this work, we have connected it to thermodynamic properties of
Lifshitz black hole.   All heat capacities of Lifshitz black holes
are positive and all their free energies are negative, which means
that all Lifshitz black holes belong to the GSBH.  These globally
stable Lifshitz black holes provide purely imaginary QNFs when
choosing a scalar perturbation, which means that it is hard  to take
the black hole out off the equilibrium.

Although we have tested a few of Lifshitz black holes, we suggest
that most of Lifshitz black holes provide purely imaginary QNMs even
if one uses  different physical field (being not a minimally coupled
scalar) as a perturbation field.

As a byproduct, we have computed area spectrum of Lifshitz black
holes by using the Maggiore's method. The constant $\gamma$ (or
$\sigma)$ of $A_n=\gamma n \hbar$ ($S_n=\sigma n \hbar)$ is given by
$\gamma=\sigma=4(\sqrt{6}-2)\pi$ for 3D Lifshitz black hole with
$G_3=1$,
 $\gamma=8\pi$ ($\sigma=\pi$) for 4D1 Lifshitz black hole,
 and  $\gamma=16\pi$ ($\sigma=2\pi$) for 4D2 Lifshitz black hole.
Even though all are not universal area spectrum, the area quantum
spectrum could be derived from QNFs of Lifshitz black holes.
Consequently, this result shows that both the horizon area and the
entropy can be quantized for the globally stable Lifshitz black
holes.

We conclude with a comment on the recent issue related to scale
covariant metric \cite{LCM} which leads to a violation of
hyperscaling of the dual field theory \cite{Huijse}. When a
hyperscaling exponent $\theta$ is zero, a scale covariant metric
reduces to a scale invariant metric of  Lifshitz metric.  It would
be interesting to answer to the question on the possible computation
of QNFs of general Lifshitz black holes with scale covariant metric.
Unfortunately, we could not obtain an analytic solution to the
Klein-Gordon equation in these backgrounds (e.g., Eq.(5.2) of the X.
Dong et al. work~\cite{LCM}).  However, according to the AdS/CFT
correspondence, it is known that the poles of the retarded Green's
function in the momentum space correspond to QNFs \cite{Birmingham}.
If one computes  such poles for the general Lifshitz black holes
using the Lif/CFT correspondence,
 it might shed some
light on finding QNFs.  This is surely beyond the scope of the
present paper, but nevertheless it is worthwhile to be explored in
future work.

\section*{Acknowledgement}

TM would like thank Y. Kwon for useful discussion.
 This work was supported by the National Research Foundation of Korea
(NRF) grant funded by the Korea government (MEST) through the Center
for Quantum Spacetime (CQUeST) of Sogang University with Grant
No.2005-0049409. Y. Myung  was partly supported by the National
Research Foundation of Korea (NRF) grant funded by the Korea
government (MEST) (No.2011-0027293).

\end{document}